\documentclass[12pt,osajnl2,preprint,showpacs]{revtex4}  
\usepackage[draft]{hyperref} 

\begin{document}

\title{Comment on "Design of a broadband highly dispersive pure silica
photonic crystal fiber" by Subbaraman \textit{et al.} }

\author{Niels Asger Mortensen}
\affiliation{MIC -- Department of Micro and Nanotechnology,
NanoDTU, Technical University of Denmark, bld. 345 east, DK-2800
Kongens Lyngby, Denmark} \email{asger@apsmail.org}

\begin{abstract}In a recent paper, Subbaraman {\it et al.} report a theoretical and
numerical study of highly dispersive pure silica photonic crystal
fiber supporting group-velocity dispersion exceeding $-2\times
10^4$~ps/nm/km. This comment argues that the authors only consider
one out of the two sides of the same coin, by not taking the
corresponding beating length into account.
\end{abstract}


\maketitle

In a recent paper, Subbaraman \textit{et al.} report a theoretical
and numerical study of highly dispersive pure silica photonic
crystal fiber~\cite{Subbaraman:2007}. The photonic crystal fiber
has an air-hole structure making the effective index profile
resemble that of a dispersion-compensating fiber with a w-profile
in the doping. As pointed out by the authors, the negative
group-velocity dispersion (GVD) originates from an
avoided-crossing of the uncoupled inner and outer modes. The
authors employ coupled-mode theory to give quite simple analytical
expressions which they confirm numerically by full-wave
simulations. While the analytical results are quite elegant and
fully correct, the authors are missing an important point in their
analysis and physical interpretations. The authors correctly show
that the maximal dispersion parameter is inversely proportional to
the coupling constant $\kappa$, see Eq. (S6),
\begin{equation}\label{eq:S6}
D_{\rm max} = \mp
\frac{\pi}{2c\kappa}\left(\frac{dn_1}{d\lambda}-\frac{dn_2}{d\lambda}\right)^2
\end{equation}
but from Eq.~(S2) it also straightforwardly follows that the
difference in propagation constant at $\omega_p$ is
\begin{equation}
\Delta\beta=B_+-B_-=2\kappa.
\end{equation}
In other words, the corresponding beating length (coupling length)
between the two supermodes is
\begin{equation}\label{eq:LB}
L_B=2\pi/\big|\Delta\beta\big|=\pi/\kappa.
\end{equation}
To put things even more clear we may rewrite Eq.~(\ref{eq:S6}) in
terms of Eq.~(\ref{eq:LB}) which gives
\begin{equation}
D_{\rm max} = \mp
\frac{L_B}{2c}\left(\frac{dn_1}{d\lambda}-\frac{dn_2}{d\lambda}\right)^2.
\end{equation}
Thus, large negative (or positive) group-velocity dispersion
inevitably comes at the price of a correspondingly long beating
length (corresponding to a small mode spacing).

As pointed out in e.g. Ref.~\cite{Mortensen:2003} this has strong
implications for the loss in general and in the present case for
the resulting group-delay of a propagating pulse. Basically, $L_B$
acts as a cut-off in the scattering rates associated with
longitudinal index variations. Only slow variations with a
characteristic length scale $\xi$ longer than $L_B$ will have a
negligible influence on the scattering while faster variations
with $\xi$ shorter than $L_B$ may course pronounced scattering
loss and intra-modal coupling. Obviously, one should in general
aim at a short $L_B$ to minimize scattering loss and intermodal
coupling effects~\cite{Love:1989,Mortensen:2003}. Longitudinal
index variations are inevitably present regardless of the
fabrication procedure. In addition to fabrication related issues
there will also be effects of microbending which has shown to
potentially be a serious source of scattering loss in large-mode
area photonic crystal fibers~\cite{Nielsen:2003}. In fact, even
macrobending will deform the index profile slightly and induce
scattering loss and intra-modal coupling.

In the present context intra-modal coupling will degrade the
effective dispersion compensating of a pulse. Imagine that the
pulse is launched in the supermode with the negative GVD, then
scattering loss will after a length of the order $\xi$ have
redistributed the power equally between the two supermodes. As
correctly shown by the authors the two supermodes have dispersion
parameter of opposite sign and thus the resulting dispersion
compensation is easily washed out and averaged to zero.

Obviously, there exists some window where intra-modal coupling of
the supermodes is negligible so that the negative GVD can be
utilized for dispersion compensation, but GVDs of the order
$-2\times 10^4$~ps/nm/km seems well outside the window. To
illustrate this it is worthwhile emphasizing existing
technology~\cite{GrunerNielsen:2000}. Practical MCVD fabricated
fibers typically operate around $-100$ to $-200$
ps/nm/km~\cite{GrunerNielsen:2000}, though more extreme values
have been reported~\cite{Auguste:2000}. These fibers have been
subject to a high degree of optimization, most likely pushing the
beating length to the maximal possible with the present
fabrication technology. The fabrication of photonic crystal fibers
is without any doubt even more challenging and thus it is
difficult to envision the realization of the GVD proposed by
Subbaraman \textit{et al.}~\cite{Subbaraman:2007}. In fact,
increasing the typical GVD by two orders of magnitude would
require their photonic crystal fiber to appear absolutely
longitudinal uniform on a 100 times longer length than in typical
state-of-the-art MCVD fabricated fibers!


\begin{thebibliography}{1}
\newcommand{\enquote}[1]{``#1''}

\bibitem{Subbaraman:2007}
H.~Subbaraman, T.~Ling, Y.~Q. Jiang, M.~Y. Chen, P.~Y. Cao, and
R.~T. Chen,
  \enquote{Design of a broadband highly dispersive pure silica photonic crystal
  fiber,} Appl. Optics \textbf{46}, 3263 -- 3268 (2007).

\bibitem{Mortensen:2003}
N.~A. Mortensen and J.~R. Folkenberg, \enquote{Low-loss criterion
and effective
  area considerations for photonic crystal fibres,} J. Opt. A: Pure Appl. Opt.
  \textbf{5}, 163 -- 167 (2003).

\bibitem{Love:1989}
J.~D. Love, \enquote{Application of a low-loss criterion to
optical wave-guides
  and devices,} IEE Proc. J \textbf{136}, 225 -- 228 (1989).

\bibitem{Nielsen:2003}
M.~D. Nielsen, N.~A. Mortensen, and J.~R. Folkenberg,
\enquote{Reduced
  microdeformation attenuation in large-mode-area photonic crystal fibers for
  visible applications,} Opt. Lett. \textbf{28}, 1645 -- 1647 (2003).

\bibitem{GrunerNielsen:2000}
L.~Gr{\"u}ner-Nielsen, S.~N. Knudsen, B.~Edvold, T.~Veng,
D.~Magnussen, C.~C.
  Larsen, and H.~Damsgaard, \enquote{Dispersion compensating fibers,} Opt.
  Fiber Technol. \textbf{6}, 164 -- 180 (2000).

\bibitem{Auguste:2000}
J.~L. Auguste, R.~Jindal, J.-M. Blondy, M.~Clapeau, J.~Marcou,
B.~Dussardier,
  G.~Monnom, D.~B. Ostrowsky, B.~P. Pal, and K.~Thyagarajan,
  \enquote{-1800ps/(nm.km) chromatic dispersion at 1.55 $\mu$m in dual concentric
  core fibre,} Electron. Lett. \textbf{36}, 1689 -- 1691 (2000).

\end{thebibliography}

\end{document}